\documentclass[a4paper,11pt]{article}
\usepackage{jinstpub} 
\usepackage{amsmath,upgreek}
\usepackage{graphicx,color}
\usepackage{multirow,rotating}

\newcommand{\Nuc}[2]{\ensuremath{^{#2}\mbox{#1}}}

\newcommand{\C}{$^{\circ}$C}

\usepackage{url} 
\usepackage{lineno}

\title{ZnO-based scintillating bolometers: New prospects to study double beta decay of $^{64}$Zn
}

\author[a]{A.~Armatol,}
\author[b]{B.~Broerman,}
\author[c]{L.~Dumoulin,}
\author[c]{A.~Giuliani,}
\author[a]{H.~Khalife,}
\author[d]{M.~Laubenstein,}
\author[c]{P.~Loaiza,}
\author[c]{P.~de~Marcillac,}
\author[c]{S.~Marnieros,}
\author[b,e,1]{S.S. Nagorny%
\note{Corresponding author.}}
\author[d]{S.~Nisi,}
\author[a]{C.~Nones,}
\author[c]{E.~Olivieri,}
\author[d,f]{L.~Pagnanini,}
\author[d]{S.~Pirro,}
\author[c,1]{D.V.~Poda,}
\author[c]{J.-A.~Scarpaci,}
\author[a]{and A.S.~Zolotarova}

\affiliation[a]{IRFU, CEA, Universit\'e Paris-Saclay, F-91191 Gif-sur-Yvette, France}
\affiliation[b]{Department of Physics, Engineering Physics and Astronomy, Queen's University, K7L 3N6, Kingston, ON, Canada}
\affiliation[c]{Universit\'e Paris-Saclay, CNRS/IN2P3, IJCLab, 91405 Orsay, France}
\affiliation[d]{INFN -- Laboratori Nazionali del Gran Sasso, I-67100, Assergi, AQ, Italy}
\affiliation[e]{Arthur B. McDonald Canadian Astroparticle Physics Research Institute, K7L 3N6, Kingston, ON, Canada}
\affiliation[f]{Gran Sasso Science Institute, I-67100, L'Aquila, AQ, Italy}

\emailAdd{sn65@queensu.ca}
\emailAdd{denys.poda@ijclab.in2p3.fr}

\abstract{The first detailed study on the performance of a ZnO-based cryogenic scintillating bolometer as a detector to search for rare processes in zinc isotopes was performed. A 7.2~g ZnO low-temperature detector, containing more than 80\% of zinc in its mass, exhibits good energy resolution of baseline noise 1.0--2.7~keV FWHM at various working temperatures resulting in a low-energy threshold for the experiment, 2.0--6.0~keV. The light yield for $\beta$/$\gamma$ events was measured as 1.5(3)~keV/MeV, while it varies for $\alpha$ particles in the range of 0.2--3.0~keV/MeV. The detector demonstrate an effective identification of the $\beta$/$\gamma$ events from  $\alpha$ events using time-properties of only heat signals. 
The radiopurity of the ZnO crystal was evaluated using the
Inductively Coupled Plasma Mass Spectrometry, an ultra-low-background High Purity Ge $\gamma$-spectrometer, and bolometric measurements. Only limits were set at the level of $\mathcal{O}$(1--100)~mBq/kg on activities of \Nuc{K}{40}, \Nuc{Cs}{137} and daughter nuclides from the U/Th natural decay chains. The total internal $\alpha$-activity was calculated to be 22(2)~mBq/kg, with a major contribution caused by 6(1)~mBq/kg of \Nuc{Th}{232} and 12(2)~mBq/kg of \Nuc{U}{234}. Limits on double beta decay (DBD) processes in \Nuc{Zn}{64} and \Nuc{Zn}{70} isotopes were set on the level of $\mathcal{O}(10^{17}$--$10^{18})$~yr for various decay modes profiting from 271~h of acquired background data in the above-ground lab. This study shows a good potential for ZnO-based scintillating bolometers to search for DBD processes of Zn isotopes, especially in \Nuc{Zn}{64}, with the most prominent spectral features at $\sim$10--20~keV, like the two neutrino double electron capture. A 10~kg-scale experiment can reach the experimental sensitivity at the level of $\mathcal{O}(10^{24})$ yr.}


\begin{document}

\maketitle
\flushbottom

\section{Introduction}\label{sec:intro}

The observation of neutrino flavor oscillations has provided evidence of the non-degenerate mass of the neutrinos and motivated a worldwide experimental effort to measure the absolute neutrino mass and the actual scheme of the neutrino mass ordering~\cite{Workman:2022ynf}. Neutrinoless double beta decay (0$\nu$-DBD) is the only practical means of determining the nature of the neutrino (Dirac or Majorana) and one of the most sensitive probes of its absolute mass~\cite{Agostini:2022,dolinski2019neutrinoless}. If observed this would imply lepton number violation, and be a direct probe for physics beyond the Standard Model. The search for 0$\nu$-DBD with different target isotopes is not only identified as a recommendation of the APPEC committee~\cite{giuliani2019double} but also provides crucial input for the theoretical modeling of nuclear matrix elements.

The recent increase of new results in the field of DBD studies is due to significant development of various detector techniques (scintillators~\cite{Abe:2023,barabash2018final,iida2016candles}, semiconductor detectors~\cite{agostini2020final,Arnquist:2023}, bolometers~\cite{Adams:2020,Brofferio:2018}, scintillating bolometers~\cite{alenkov2019first,Azzolini:2022,augier2022final}, time-projection chambers~\cite{anton2019search,xenon2019observation,Si:2022,novella2022measurement}, and tracking calorimeters~\cite{arnold2015results,waters2017latest}), the establishment of highly effective deep material purification, and last but not least, the development of high quality crystals with embedded and highly-enriched isotopes of interest (e.g. CaF$_2$, high purity \Nuc{Ge}{76}, TeO$_2$, Zn\Nuc{Se}{82}, \Nuc{Cd}{106}WO$_4$, \Nuc{Cd}{116}WO$_4$, Li$_2$\Nuc{Mo}{100}O$_4$, Ca\Nuc{Mo}{100}O$_4$, Zn\Nuc{Mo}{100}O$_4$) \cite{danevich2017radiopure,danevich2018radioactive,Barabash:2020a}. 
These advances, however, have lead to focusing experimental efforts on a short-list of DBD-active isotopes, such as \Nuc{Ca}{48}, \Nuc{Ge}{76}, \Nuc{Se}{82}, \Nuc{Mo}{100}, \Nuc{Cd}{116}, \Nuc{Te}{130}, \Nuc{Nd}{150} and \Nuc{Xe}{136} \cite{Agostini:2022,Pritychenko:2023,Barabash:2020,Belli:2020a}.
Other isotopes\footnote{There are 69~potentially DBD-active natural isotopes~\cite{tretyak2002tables}.} are less studied for reasons specific to each individual isotope. Further details on techniques in the search for rare decays can be found in~\cite{Agostini:2022,d2021neutrinoless,zolotarova2021bolometric,danevich2018radioactive,danevich2017radiopure}.

Zinc contains two potentially DBD-active natural isotopes, namely \Nuc{Zn}{64} and \Nuc{Zn}{70}, whose properties are listed in table~\ref{tab:zn}. \Nuc{Zn}{64} is one of a few DBD-active nuclei to have a high natural abundance ($\sim$48\%) capable of use in a large scale experiment without expensive isotopic enrichment. Furthermore, the relatively high transition energy of \Nuc{Zn}{64} ($Q_{2\beta}$~=~1096~keV) makes it energetically allowed for both double electron capture (2$\varepsilon$) and electron capture with positron emission ($\varepsilon\beta^+$) channels of DBD~\cite{audi2003ame2003}. There is a strong motivation to search for these processes as it could clarify the contribution of the right-handed current admixture in weak interactions~\cite{Hirsch1994,fukuyama2022neutrinoless}. Despite several detector materials and techniques currently available to study DBD processes in Zn isotopes, all of them have some drawbacks. A survey of these techniques follows and is given in table~\ref{tab:comparison} in order of increasing zinc mass fraction.

\begin{table}[h!]
\caption{Transition, energy released, isotopic abundance, and decay modes for potentially DBD-active natural zinc isotopes.}
\begin{center}
\begin{tabular}{cccc}
\hline
Transition & Energy release & Isotopic abundance & Decay modes \\ 
~ & $Q_{2\beta}$ [keV]~\cite{wang2021ame} & [\%]~\cite{meija2016isotopic} & ~ \\ 
\hline
\Nuc{Zn}{64} $\rightarrow$ \Nuc{Ni}{64} & 1095.7(0.7) & 49.17(75) & $2\varepsilon$, $\varepsilon\beta^+$ \\
\Nuc{Zn}{70} $\rightarrow$ \Nuc{Ge}{70} & 998.5(2.2) & 0.61(10) & $2\beta^-$ \\
\hline
\end{tabular}
\end{center}
 \label{tab:zn}
\end{table}

ZnWO$_4$ scintillating crystals contain zinc at the level of 21\% in mass. This naturally radiopure scintillating material was proposed as a detector for DBD studies of Zn and W isotopes and to search for dark matter particles in 2005~\cite{danevich2005znwo4}. Later, the technology of large volume crystals has been well-developed~\cite{nagornaya2008growth,nagornaya2009large}, as well as an improvement of its radiopurity was achieved through the multi-stage crystallization process~\cite{barabash2016improvement}. The best limits on DBD processes in Zn isotopes  achieved in low-background long-term measurements with ZnWO$_4$ scintillating crystals are of $\mathcal{O}(10^{19}$--$10^{20})$ yr. In should be emphasized that both the relatively poor energy resolution (about 9\% for 662~keV $\gamma$~quanta) and the low Zn-mass fraction cause a reduced experimental sensitivity.	

Despite the excellent performance of ZnWO$_4$ also as Transition Edge Sensor (TES)-carriers~\cite{bavykina2008investigation} developed and used in the R\&D program of the CRESST experiment and its high radiopurity~\cite{belli2011radioactive}, there are no reported long-term low-background cryogenic measurements with a large volume ZnWO$_4$ crystal acting as a scintillating bolometer. However, excellent results with a 1~cm$^3$ ZnWO$_4$ scintillating bolometer have been recently achieved~\cite{dumoulin2022assessment}. As it was shown with many other crystals~\cite{poda2021scintillation}, a simultaneous record of a scintillation pulse and a phonon signal allows for an effective particle identification leading to a significant background reduction. The bolometric technique also typically provides an excellent energy resolution ($\sim$0.1\%) in the phonon channel~\cite{poda2017low}. Both features lead to the enhancement of the experimental sensitivity over purely-scintillating ZnWO$_4$.

\begin{table}[h!]
\caption{Survey of Zn-containing detector techniques. Listed with the target material is the fraction of Zn by weight, detector type, and major drawbacks to the method. (N.D. = not determined, T.B.A. = to be analysed, Scint. Bol. = scintillating bolomter, HPGe = High Purity Ge $\gamma$-spectrometer.)}
\begin{center}
\begin{tabular}{lcccc}
\hline
Target      & Fraction of       & Detector          & Major & Reference \\  
material &  Zn [wt \%] & type & drawback(s) & ~ \\ \hline
\multirow{2}{*}{ZnWO$_4$} & \multirow{2}{*}{21} & Scintillator & Poor FWHM & \cite{belli2011final} \\ \cline{3-5}
~ & ~ & Scint. Bol. & N. D. &  \cite{bavykina2008investigation} \\ \hline
Li$_2$Zn$_2$(MoO$_4)_3$      & 21              & Scint. Bol. & \parbox{2.7cm}{\centering Interference with DBD of \Nuc{Mo}{100}; \\Low radiopurity} & \cite{bashmakova2009li2zn2}\\ \hline
CdZnTe	    & 21           & Semiconductor     & \parbox{3.5cm}{\centering \vspace{0.1cm}Interference with DBD of \Nuc{Cd}{116},\Nuc{Te}{130}; \\Low radiopurity\vspace{0.1cm}} & \cite{ebert2016results} \\  \hline
ZnMoO$_4$   & 29           &	Scint. Bol. & \parbox{2.7cm}{\centering \vspace{0.1cm}Interference with DBD of \Nuc{Mo}{100};\\ Low light yield\vspace{0.1cm}} & \cite{beeman2012performances} \\ \hline
\multirow{2}{*}{ZnSe}        & \multirow{2}{*}{44}           & \multirow{2}{*}{Scint. Bol.}  & Interference with  & \multirow{2}{*}{\cite{azzolini2020search}} \\ 
~ & ~ & ~ & DBD of \Nuc{Se}{82} & \\ \hline
ZnO	        & 80           & Scint. Bol. &	\parbox{2.5cm}{\centering T.B.A.} &  This work \\ \hline
\multirow{3}{*}{Zn metal}	&  \multirow{3}{*}{100}            & \multirow{2}{*}{HPGe}	             & Poor detection  & \multirow{2}{*}{\cite{bellini2021search}} \\ 
~ & ~ & ~ & efficiency & \\ \cline{3-5}
~ & ~ & Superconductor & R\&D, T.B.A. &  \cite{Augier:2021} \\ \hline
\end{tabular}
\end{center}
 \label{tab:comparison}
\end{table}

Another Zn-containing scintillating crystal, namely Li$_2$Zn$_2$(MoO$_4)_3$, has been established in 2009 and tested as a potential target material for 0$\nu$-DBD searches of \Nuc{Mo}{100} ($Q_{2\beta}$ = 3034~keV) acting as a scintillating bolometer~\cite{bashmakova2009li2zn2}. Containing Zn at the level of 21\% in mass, this crystal could be also used to search for DBD process occurring in Zn isotopes. However, a very poor yield of scintillation light observed in the first cryogenic test makes it impossible to achieve all typical features of the scintillating bolometer technique though simultaneous record of scintillation light and phonon signal. Moreover, only small volume crystals (less than 0.2~kg in mass) could be produced at that time. The combination of technological issues during the crystal growth and its poor performance as a scintillating bolometer denied further studies with this compound.

A CdZnTe (CZT) semiconductor compound, containing 21\% of Zn in mass, is a very promising radiation detector due to its excellent energy resolution (FWHM = 1\% for 662~keV $\gamma$~quanta). However, it also contains DBD-active \Nuc{Cd}{116} ($Q_{2\beta^-} = 2814$~keV) and \Nuc{Te}{130} ($Q_{2\beta^-} = 2529$~keV) isotopes with larger transition energies and shorter half-lives, which are responsible for an irreducible background for DBD processes of Zn isotopes. Moreover, the long-lived beta-active \Nuc{Cd}{113} isotope ($Q_{\beta} = 323.8$~keV, $T_{1/2}=8\times10^{15}$~yr~\cite{wang2021ame,belli2007investigation}), present in natural Cd at the level of 12\%, is a main background component at low energies below $\sim$325~keV, preventing the study of DBD processes in Zn isotopes with signatures at low energies. From the technological point of view, CZT crystals larger than $2\times2\times2$~cm are not available yet. The best limits achieved in the framework of COBRA experiment with CZT semiconductors regarding DBD processes of Zn isotopes are of $\mathcal{O}(10^{18})$~yr.

From 2010, ZnMoO$_4$ crystals were considered as the most promising target material for 0$\nu$-DBD studies of \Nuc{Mo}{100}, due to their excellent performance as a scintillating bolometer in terms of energy resolution in phonon channel and high radiopurity~\cite{beeman2012performances, cardani2014first,armengaud2015development}. Moreover, the possibility to perform particle identification only through the signal properties in phonon channel~\cite{beeman2012performances} was demonstrated in ZnMoO$_4$ for the first time among all previously tested crystals as a cryogenic bolometer. This is a useful feature for future large-scale experiments, since it allows to minimize the total number of electronics channels, acquiring only the heat channel, while not compromising the experimental sensitivity. Thanks to the extensive R\&D program, radiopure crystals up to 2~kg in mass were produced from natural molybdenum~\cite{berge2014purification} and enriched in \Nuc{Mo}{100}~\cite{barabash2014enriched}. A ZnMoO$_4$ compound contains approximately 29\% of zinc; however, the presence of another DBD-active isotope, i.e.~\Nuc{Mo}{100}, with a shorter half-life value and larger decay energy makes it difficult to study DBD processes in Zn.

Recently, experimental data acquired with Zn\Nuc{Se}{82} scintillating bolometers ($\sim$45\% Zn by mass) within the CUPID-0 experiment were analyzed to set new limits on several modes of DBD processes in \Nuc{Zn}{64} and \Nuc{Zn}{70} isotopes up to $\mathcal{O}(10^{21}$--$10^{22})$~yr~\cite{azzolini2020search}. Both types of ZnSe crystals, produced from natural Se and enriched in \Nuc{Se}{82}, worked well as scintillating bolometers and exhibited decent performance (FWHM $\approx30$~keV in phonon channel at 3~MeV), excellent pulse-shape discrimination ability and high radiopurity~\cite{beeman2013performances,azzolini2019background}. These new limits were obtained from data collected with 22 enriched crystals and one natural crystal with 9.18 kg of ZnSe active mass for a 11.34 kg$\times$yr of total collected exposure. The experimental sensitivity was limited by the presence of a high amount of DBD-active \Nuc{Se}{82} isotope ($Q_{2\beta^-}=$ 2998~keV), responsible for the major background contribution, and by a limited acquisition time. It should be emphasized that from a technological point of view, ZnSe crystal growth is a very complicated process and not well-established yet. To date, ZnSe crystals with dimensions up $\varnothing 45\times55$~mm could be produced by the High-Pressure Bridgman-Stockbarger technique with a yield of ``ready-to-use'' crystals less than 50\%~\cite{sergeSe}. Therefore, the use of ZnSe crystals in further studies of DBD processes in Zn isotopes is unfavorable.

Complementary studies were performed with 10~kg of a highly purified Zn metal measured on the ultra-low-background (ULB) HPGe $\gamma$-spectrometer at the Laboratori Nazionali del Gran Sasso of the INFN (LNGS, Italy) over 828~h~\cite{bellini2021search}. Through the optimization of the Zn sample geometry and high radiopurity of the Zn metal, the highest limits on some modes of DBD processes were established at the $\mathcal{O}(10^{21})$ yr. At the same time, the experimental sensitivity cannot be further significantly improved due to the limitation of detection efficiency in this ``source $\neq$ detector'' approach; there is also a limit to the effective mass that could be placed around HPGe detector that does not deteriorate the optimal detection efficiency. Moreover, with only emitted $\gamma$'s being detected, this technique unable to distinguish between $0\nu$ and $2\nu$ modes of $\varepsilon\beta^+$ decay of \Nuc{Zn}{64} and cannot be used for studies of $2\beta^-$ decay of \Nuc{Zn}{70}. It should be noted that measurements with Zn metal samples on HPGe detectors were one of the first techniques to study DBD processes in Zn~\cite{norman1985improved,kim2007searches}. A low efficiency of the passive source technique can potentially be overcome by using metallic zinc as a superconducting absorber with a TES phonon readout; this innovative low-threshold detector technology is under development by Ricochet collaboration for the detection of coherent elastic neutrino-nucleus scattering \cite{Augier:2021,Chen:2022}.

As one can see, none of the target materials and experimental approaches listed above are optimal to search for DBD processes in Zn isotopes. To reach the highest sensitivity, a detector should fulfill several requirements:
\begin{itemize}
\item[a)] possess a high Zn-content;
\item[b)] elements embedded in its chemical formula should be light in mass;
\item[c)] chemical formula should be free from other DBD-active elements;
\item[d)] possess a high radiopurity;
\item[e)] have a well-developed technology for detector production;
\item[f)] work as a cryogenic scintillating bolometer.
\end{itemize}

We propose to use a ZnO-based scintillating bolometer, with the highest zinc mass fraction (more than 80\%), to search for DBD processes in Zn isotopes. ZnO crystals have never been used before as radiation detectors, while typically utilized as piezoelectric crystals, wafers for powerful LEDs in the blue and UV spectral ranges, and also find use in gas sensors, varistors, and generators of
surface acoustic waves~\cite{nickel2006zinc}. Oxygen, the only remaining element in the chemical formula, is not a DBD-active element, and will not contribute to the detector background. Here we present results on the first cryogenic test of a ZnO-based scintillating bolometer, its production, evaluation of performance and radiopurity, and studies of prospects for DBD searches.

\section{Crystal production}\label{sec:production}

ZnO (zincite) single crystals belong to the structural type of wurtzite with a hexagonal cell that have the space symmetry group (P63mc) with the lattice constants $a = 3.250$~\AA, $c = 5.207$~\AA. Crystals demonstrate a very strong  anisotropy of their physical properties. Zincite crystals considered to be a heavily doped weakly compensated semiconductor material of the A$^2$B$^6$ type, in which intrinsic defects (interstitial zinc) and, to a lesser extent, oxygen vacancies serve as donors (n-type conductivity). Some others characteristics of ZnO crystals are as follows~\cite{rodnyi2011optical}: the melting point is 1975\C, the density is 5.67~g/cm$^3$, and the band gap is 3.44~eV (1.6~K) and 3.37~eV (300~K).

The ZnO crystal used in the present study was grown by the hydrothermal method~\cite{kortunova2008hydrothermal}, the most efficient technique for large batches of crystals with similar properties. To produce large-volume ZnO single crystals, this method has been adapted for autoclaves with 80-mm-diameter containers that allows to grow up to 20 large single crystals ($50\times50\times12$ mm each) simultaneously.

Special attention was paid to prepare the starting charge and seed materials. A high-purity ZnO powder was pressed into tablets and annealed at 1000\C~under air atmosphere. These cylindrical tablets were then used as a starting charge. Seeding plates were cut from previously-grown ZnO single crystals oriented along the [0001] crystallographic direction. Then seed plates were polished and etched prior the growth process.

ZnO crystal growth was carried out by the direct temperature drop in alkaline solutions (KOH + LiOH + NH$_4$OH) at the crystallization temperature of approximately 330--360\C. The temperature drop between crystallization and dissolution zone was approximately 8--15\C, and under pressure of 30--40 MPa (more details can be found in~\cite{kortunova2008hydrothermal}). The growth rate in the [0001] direction was about 0.12~mm/day. A small ZnO sample ($10.6\times11.0\times11.0$~mm) used in this study was cut from a large crystal ($50\times50\times12$ mm) produced following this technique. Then two opposite faces were optically polished, while the lateral surface was ground to enhance light output.

The concentration of the most common radioactive elements in the ZnO crystal was measured by an Inductively Coupled Plasma Mass Spectrometry analysis (ICP-MS, Agilent Technologies model 7500a) at LNGS. The analysis was performed in a semiquantitative mode; the instrument was calibrated with a single standard solution containing 10 ppb of Li, Y, Ce and Tl. The uncertainty is approximately 25\% of given concentration values, listed in table~\ref{tab:icpms}.

\begin{table}[h!]
\caption{The concentration of the most common natural radioactive elements (and their activity) in the ZnO crystal measured by an ICP-MS instrument. The uncertainty is approximately 25\% of given concentration values.
} 
\begin{center}
\begin{tabular}{llll}
\hline
Element & Concentration & \multicolumn{2}{c}{Specific activity}  \\ 
~ & ~ [ppb] & [mBq/kg] & Nuclide \\ 
\hline
K       & $\leq 10000$ & $\leq 310$ & $^{40}$K \\
Th      & $\leq 1$     & $\leq 4$   & $^{232}$Th \\
U       & $\leq 1$     & $\leq 12$  & $^{238}$U \\
\hline
\end{tabular}
\end{center}
 \label{tab:icpms}
\end{table}

Being produced from the raw materials and reagents of 99.98\% chemical purity grade that are potentially contaminated with natural radioactive nuclides, the ZnO crystal exhibits rather high chemical purity with respect to U/Th and K content, which could be a result of an effective impurity segregation during the crystal growth process.

\section{HPGe measurements}\label{sec:hpge}

In order to improve sensitivity to possible radioactive contaminats, the ZnO sample was measured for 1107~h with an ULB~HPGe $\gamma$-spectrometer in the STELLA (SubTerranenan Low Level Assay) facility at LNGS~\cite{laubenstein2017screening}. Thanks to the deep underground location of the STELLA facility (corresponding to more than 3600~m~water equivalent of overburden), the reduction of muon flux is a factor $10^6$. The ULB HPGe detector has a volume of 468~cm$^3$ and an energy resolution of 1.8~keV at 1332~keV. The passive shield of the detector consists of low-radioactivity lead ($\sim$25~cm), copper ($\sim$5~cm), and ancient lead ($\sim$2~cm) on the inner part of the copper shield. The set-up is sealed in an air-tight stainless steel box continuously flushed with a high purity nitrogen gas to avoid the presence of residual environmental radon.

\begin{figure}[h!]
\centering
\includegraphics[width=0.8\textwidth]{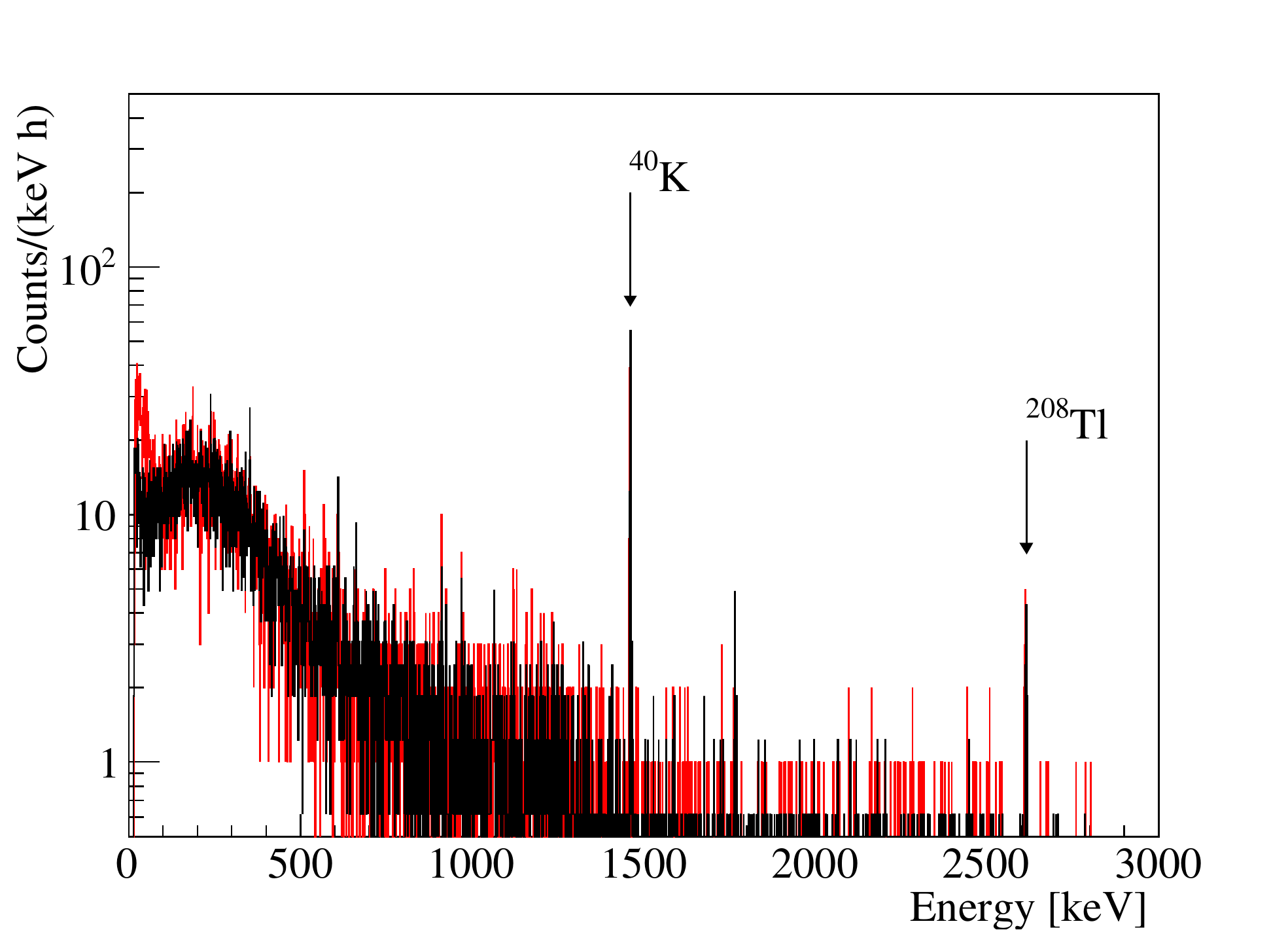}
\caption{Energy spectra of background (black) and the 7.2~g ZnO crystal sample (red), measured using the ULB HPGe detector at underground laboratory (LNGS) over 1798~h and 1107~h, respectively. Spectra are normalized to the acquisition time of the ZnO sample measurement. The normalized spectra are almost indistinguishable.}
\label{fig:hpge}
\end{figure}

The measured energy spectra of the background (with no sample) and of the ZnO crystal sample are shown in figure~\ref{fig:hpge}, normalized to the acquizition time of the ZnO sample. The normalized spectra are almost indistinguishable in the wide energy range, except for a low-energy region (below 100~keV), where some excess of counts is observed for ZnO crystal sample. The detection efficiencies were obtained using Monte-Carlo (MC) simulation in the MaGe framework~\cite{boswell2011mage} based on the GEANT4. We found no evidence of the $\gamma$ peaks, which can be definitely ascribed to decays of natural radionuclides in the ZnO sample. Therefore, we set only limits on specific activities using the Feldman-Cousins method~\cite{feldman1998unified}. The results are presented in table~\ref{tab:hpge}; upper limits on activities of U/Th radionuclides, as well as \Nuc{K}{40} from natural radioactivity or \Nuc{Cs}{137} of anthropogenic origin are set on the level of $\mathcal{O}$(1--100)~mBq/kg.

The established limit on activity of \Nuc{K}{40} in the ZnO sample, less than 220~mBq/kg, corresponds to the upper limit of the natural potassium contamination in the crystal less than 7~ppm. This value is in agreement and slightly improves the limit on the potassium concentration obtained in ICP-MS measurements, less than 10~ppm (see section~2).

\begin{table}[h!]
\caption{Internal radioactive contamination of the 7.2~g ZnO crystal sample measured over 1107~h using the ULB~HPGe detector at LNGS. The upper limits are given at 95$\%$~C.L.
} 
\begin{center}
\begin{tabular}{lll}
\hline
Chain           & Radionuclide           & Activity [mBq/kg] \\ \hline
                & \Nuc{K}{40}       & $\leq 220$   \\
                & \Nuc{Cs}{137}     & $\leq 5$     \\ \hline
\multirow{2}{*}{\Nuc{Th}{232}}   & \Nuc{Ra}{228}     & $\leq 16$   \\
                & \Nuc{Th}{228}     & $\leq 23$   \\ \hline
\Nuc{U}{235}    & \Nuc{U}{235}      & $\leq 64$   \\ \hline             
                & \Nuc{Th}{234}     & $\leq 470$    \\
\Nuc{U}{238}    & \Nuc{Pa}{234\rm{m}}& $\leq 340$   \\ 
                 & \Nuc{Ra}{226}     & $\leq 12$  \\\hline
\end{tabular}
\end{center}
\label{tab:hpge}
\end{table}

\section{Low-temperature measurements}\label{sec:cryo}

\begin{figure}[h!]
\centering
\includegraphics[width=0.8\textwidth]{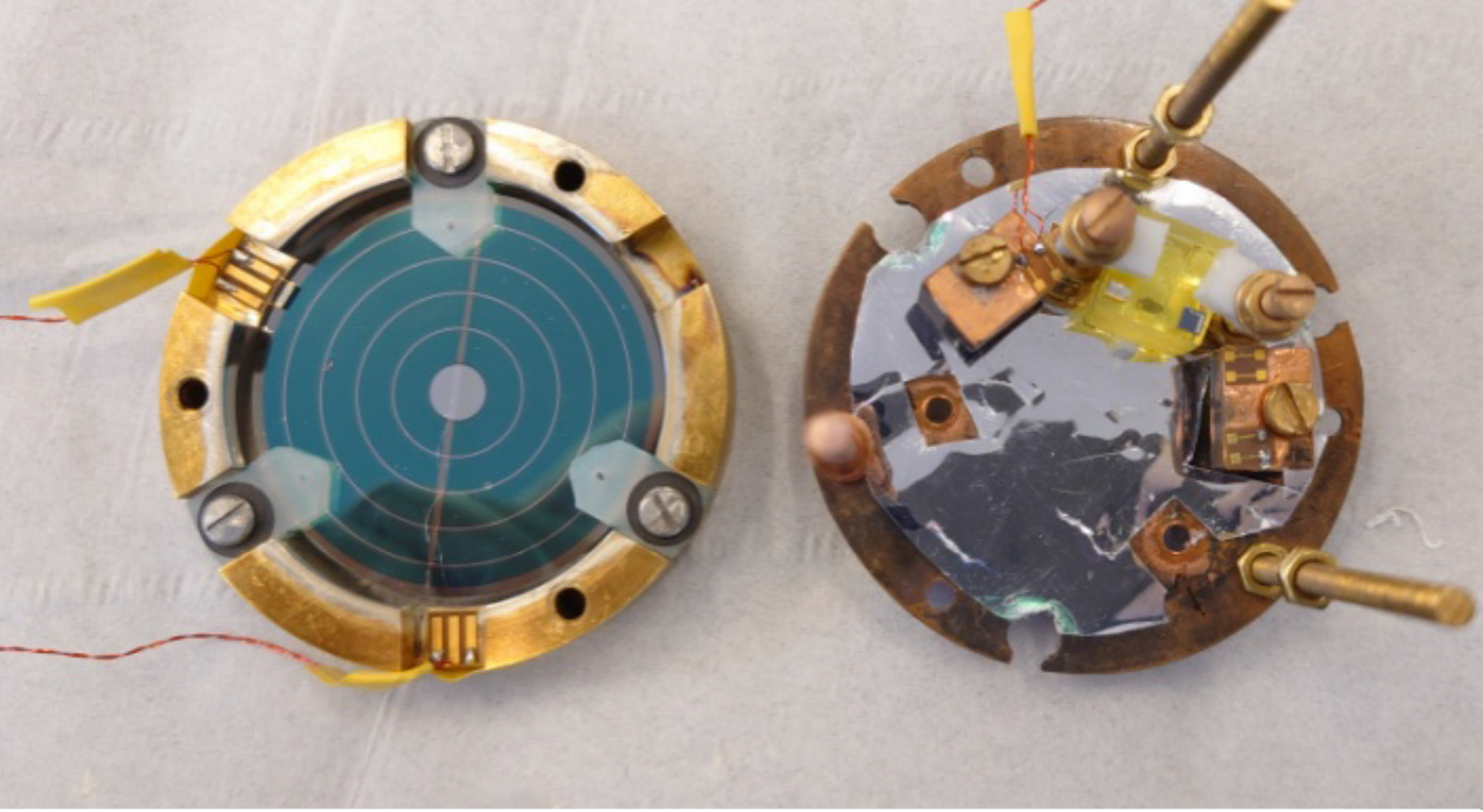}
\caption{A bolometric light detector (Left), based on a Ge disk with a size of $\varnothing 44\times0.17$ mm and instrumented with a $\sim$5~mg NTD Ge thermistor, and a ZnO scintillating bolometer (Right), based on a $10.6\times11\times11$ mm crystal and equipped with a $\sim$5~mg NTD Ge and with a Si:P heater. Both absorbers are mounted on the Cu holder using either PTFE supporting elements (ZnO) or Al$_2$O$_3$ balls and plastic clamps. The Ge LD was then placed on the top of the ZnO detector assembly.}
\label{fig:bolometer}
\end{figure}

A composite bolometric detector aiming at measuring simultaneously heat and scintillation of the ZnO sample at millikelvin temperatures was made of two modules with the mounted ZnO crystal and Ge wafer, shown in figure~\ref{fig:bolometer}.

The ZnO sample was equipped with a neutron-transmutation-doped (NTD) Ge thermistor~\cite{haller1994advanced} (with a dimension of $2.0\times1.5\times0.3$~mm), a heat-to-voltage transducer, to register the temperature variation induced by interacting particles in the detector media. The NTD Ge sensor was glued directly on crystal surface using a two-component epoxy glue (Araldite\textregistered). In addition, a P-doped silicon heater~\cite{andreotti2012production} was glued on the same crystal face aiming at injecting periodically a constant power (mimicking monoenergetic particle signals used for the off-line stabilization of the detector thermal response fluctuations~\cite{alessandrello1998methods}). Ultrasonic wire-bonding was used to connect the golden pads of the NTD Ge thermistor to the gold-plated-on-Kapton contacts glued on the copper housing with two Au wires ($\varnothing 25~\upmu$m), providing both thermal and electrical links, while the heater was bonded with two Al wires ($\varnothing25~\upmu$m). The crystal was mounted on a copper plate using PTFE (polytetrafluoroethylene) clamps and brass screws. The entire internal side of the bottom copper plate of the detectors module and the lateral side of the Cu housing were covered by a reflective film (Vikuiti\texttrademark) to improve light collection on a photodetector.

A bolometric light detector (LD) was made of a high-purity Ge wafer ($\varnothing44\times0.17$~mm), equipped with a small NTD Ge sensor ($2.2\times0.8\times0.6$~mm), which was mounted with the help of Al$_2$O$_3$ balls ($\varnothing 1.5$~mm) and polypropylene supporting elements in a Cu housing. In order to decrease the reflectivity of the wafer, it was coated by a 70-nm SiO layer~\cite{mancuso2014experimental}. Moreover, several concentric Al electrodes (100-nm-thick, 3.8-mm-pitch) have been deposited on one side of the Ge wafer to exploit heat signal amplification in the presence of the electric field, known as the Neganov-Trofimov-Luke (NTL) effect~\cite{neganov,luke1988voltage}. More details on construction, operation, and characterization of such type of bolometric LDs can be found in~\cite{novati2019charge} (the detector used in the present work is labeled as NTLLD2 there).

We operated the ZnO scintillating bolometer in the aboveground cryogenic laboratory of the IJCLab (Orsay, France), using a pulse-tube-based \Nuc{He}{3}/\Nuc{He}{4} dilution refrigerator~\cite{mancuso2014aboveground}. The assembled detector module was installed in the cryostat on the copper plate mechanically decoupled from the mixing chamber by three springs to reduce vibrations induced by the pulse-tube of the cryostat. The outer vacuum chamber of the cryostat is surrounded by a 10-cm-thick lead to suppress the environmental $\gamma$ background. Such shielding also helps to mitigate the pile-up problem of large-volume (tens of cm$^3$) thermal detectors with a typical response of NTD-based sensors in the millisecond–second range. Taking into account that we used rather small ZnO (1.3~cm$^3$) and Ge (0.3~cm$^3$) absorbers, we do not expect a significant impact from pile-up on the bolometric characterization of the composite detector. 

We realized two cryogenic runs (labeled as Run I and II), both at 15~mK temperature stabilized on the detector plate. In the first run, a \Nuc{U}{238}/\Nuc{U}{234} $\alpha$ source was placed inside the ZnO detector housing, while the second cryogenic run was performed without the calibration source. In addition, two TeO$_2$ bolometers of 4 cm$^3$ each were faced to another side of the Ge LD in Run I, extending possibilities of the LD calibration as detailed below.

The detector channels were readout using a room temperature low-noise electronics based on DC-coupled voltage sensitive amplifiers~\cite{arnaboldi2002programmable}. The NTDs were biased with $\sim$1--5~nA currents through a few G$\Omega$ load resistances. The NTD resistances were measured in the range of hundreds k$\Omega$ to a few M$\Omega$, depending on the choice of working points. In Run I, the working points were set aiming at maximizing detectors' voltage signal per unit of energy; the optimization was done by scanning signals induced by heater (for ZnO) and LED (for Ge LD) depending on the NTDs currents. In Run II, we polarized detectors stronger to reduce their sensitivity to fluctuations/instabilities of the thermal bath temperature and to mitigate non-linearity in the detectors' response. The continuous stream data were acquired by 16-bit ADC (National Instruments NI USB-6218 BNC) with a 5~kHz sampling rate; a Bessel cut-off frequency was set at 675~Hz. A total duration of detectors' low-temperature characterization was about two and three weeks in Run I and II respectively.

\section{ZnO scintillating bolometer performance}\label{sec:performance}

The detector performance was evaluated by analysis of the acquired data processed using an optimum filter technique~\cite{gatti1986processing}, realized with the help of a MATLAB-based application. A half of second window was chosen for the data processing of both channels (LD and ZnO), as a good compromise between the time response of the detectors, their counting rates, and a window length for a more proper investigation of low frequency noise figure. A signal template of the ZnO bolometer was made by summing up signals with energies $\sim$0.2--2.0~MeV, while signals with an order of magnitude lower energies, mainly muon-induced events, were used for the construction of the LD average pulse. The noise template of both detectors is based on 10000~individual noise waveforms of corresponding channels. Events were triggered in the filtered data with energies above a threshold of $5~\times$~RMS noise. Then for each triggered event we use a signal maximum at the filter output as an energy estimate and we compute several parameters to ascribe a pulse shape. Parameters relevant for the present study are following:
\begin{itemize}
\item[a)] Rise time -- a signal rising part from 10\% to 90\% of signal amplitude;
\item[b)] Decay time -- a signal descending part from 90\% to 30\% of signal amplitude;
\item[c)] Correlation -- a Pearson's linear correlation coefficient between a triggered signal and an average pulse, both after the filtering;
\item[d)] PSD parameter defined as a ratio of a fitted amplitude (template vs.~particle signal) to a filtered amplitude (primary energy estimate); more details about this parameter can be found in~\cite{bandac20200nu2beta}.	
\end{itemize}

Pulse-shape parameters, such as Rise time and Decay time listed in table~\ref{tab:performance}, show that both bolometers have a comparatively fast response for NTD-instrumented thermal detectors. This observation can be explained by the small size of the absorbers and their thermistors, and it shows that ZnO, in addition to the widely-used Ge, has good material properties for low-temperature applications. As it was expected, a stronger polarization of NTDs in Run II resulted in a faster time response than that of Run I.

\begin{figure}[h!]
\centering
\includegraphics[width=0.7\textwidth]{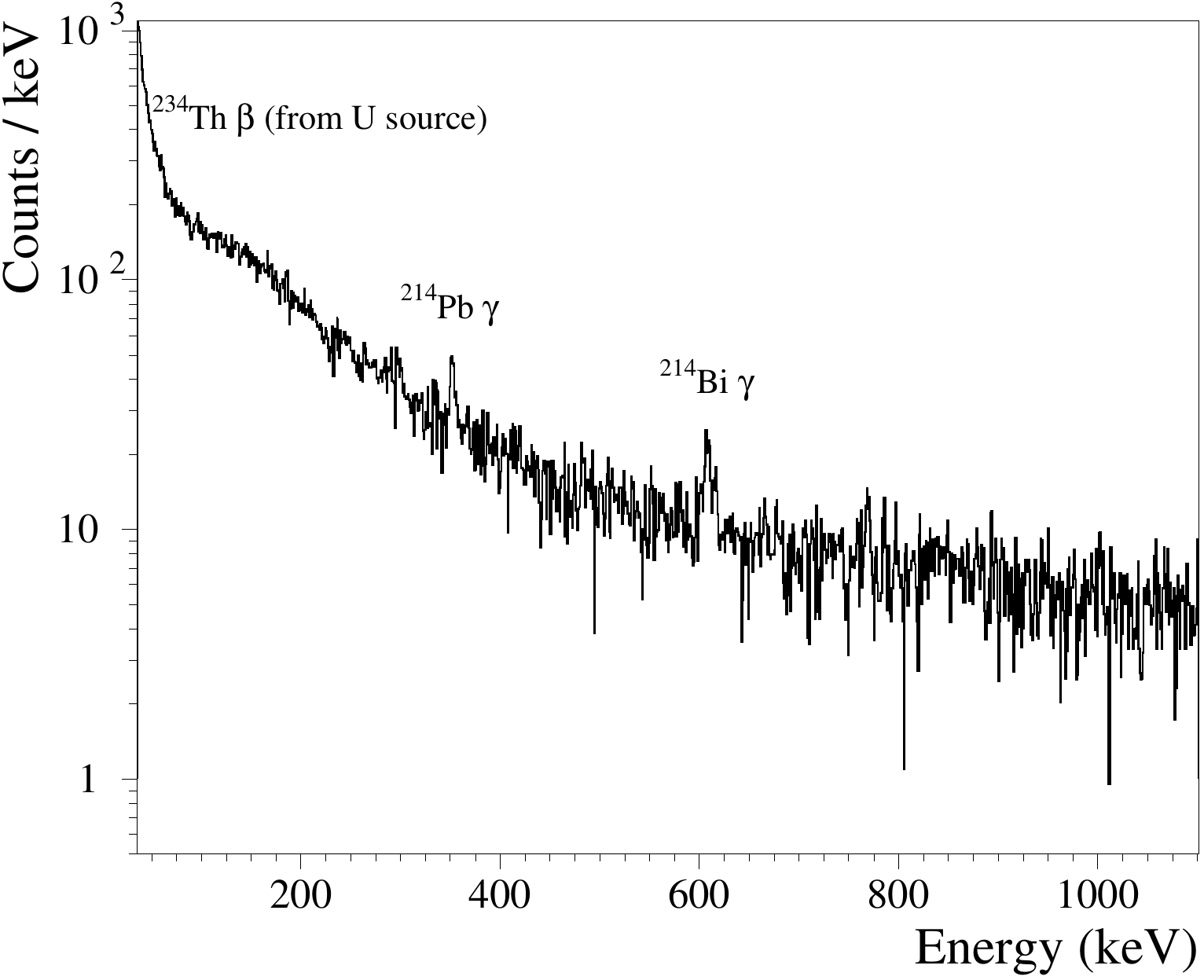}
\caption{Energy spectrum of events registered with the 7.2~g ZnO bolometer in 96-h-long low-temperature measurements (Run I) in a cryostat at sea level. The $\gamma$ peaks of \Nuc{Pb}{214} and \Nuc{Bi}{214} observed in the spectrum are originated to environmental radioactivity, while the low-energy region is dominated by beta decays of \Nuc{Th}{234}, present in the decay chain of a \Nuc{U}{238}/\Nuc{U}{234} $\alpha$ source used to irradiate the ZnO detector. }
\label{fig:spectrum}
\end{figure}

Amplitude distributions of ZnO-detected events in both Runs contain several $\gamma$ peaks of \Nuc{Pb}{214} and \Nuc{Bi}{214} from environmental radioactivity, as illustrated in figure~\ref{fig:spectrum}. It allows us to calibrate the ZnO bolometer and thus to characterize its performance in terms of the voltage signal amplitude per unit of the deposited energy (in $\upmu$V/keV) and the energy resolution of the noise baseline at the filter output (in keV FWHM), both reported in table~\ref{tab:performance}. The ZnO bolometer's sensitivity in Run I was about 0.3~$\upmu$V/keV, which is a comparatively high value for macrobolometers. Despite of a higher working temperature of NTD in Run II, the ZnO bolometric signal is found to be still notable, around 0.09~$\upmu$V/keV. The ZnO baseline noise in Run I, 1~keV FWHM, is among the best ever achieved with single-crystals-based bolometers with sizes of a few cm$^3$ or larger being tested in this pulse-tube-operated cryostat~\cite{tenconi2015development,mancuso2016development,novati2018sensitivity,khalife2021cross}. In Run II, the ZnO bolometer is characterized by a factor of 3~higher noise level (2.7~keV FWHM), in correlation with the drop of the sensitivity. The difference in the baseline noise is not crucial for the ZnO bolometer energy resolution, which is found to be relatively high, especially if compared to room temperature scintillating detectors.

\begin{table}[h!]
\caption{Performance of ZnO and Ge channels of the composite bolometric detector operated at 15 mK. The difference in the working points of the bolometers in Runs I and II (represented by the NTD resistances) is responsible for the different detector performance. Results achieved with the LD being in the NTL amplification mode (60~V electrode bias) are labeled with *.} 
\begin{center}
\begin{tabular}{l|ll|ll}
\hline
Detector    &   \multicolumn{2}{c|}{ZnO} & \multicolumn{2}{c}{Ge LD} \\
Bolometric measurements & Run I & Run II    & Run I & Run II \\ \hline
NTD working resistance [M$\Omega$]  & 2.5   & 0.42  & 4.2   & 0.54 \\
& & & & \\
Rise time [ms]	& 2.45    &	0.84  & 1.59 (1.30$^*$) & 0.95 \\
& & & & \\
Decay time [ms] & 15.3   & 5.0   & 10.4 (10.8$^*$)    &	5.5 \\ 
& & & & \\
\parbox{5cm}{Signal amplitude per unit of\\ deposited energy [$\upmu$V/keV]} &  0.33  & 0.09 & 3.2 (38$^*$)   & 0.99 \\ 
& & & & \\
\parbox{5cm}{Energy resolution [keV FWHM] of baseline noise} &    1.0 & 2.7   & 0.29 (0.04$^*$)   & 0.24 \\ 
& & & & \\
\parbox{5cm}{Energy resolution [keV FWHM] at 352 keV}  & 4.9 & 8.8 & - &	- \\
\hline
\end{tabular}
\end{center}
\label{tab:performance}
\end{table}

\begin{figure}[h!]
\centering
\includegraphics[width=0.7\textwidth]{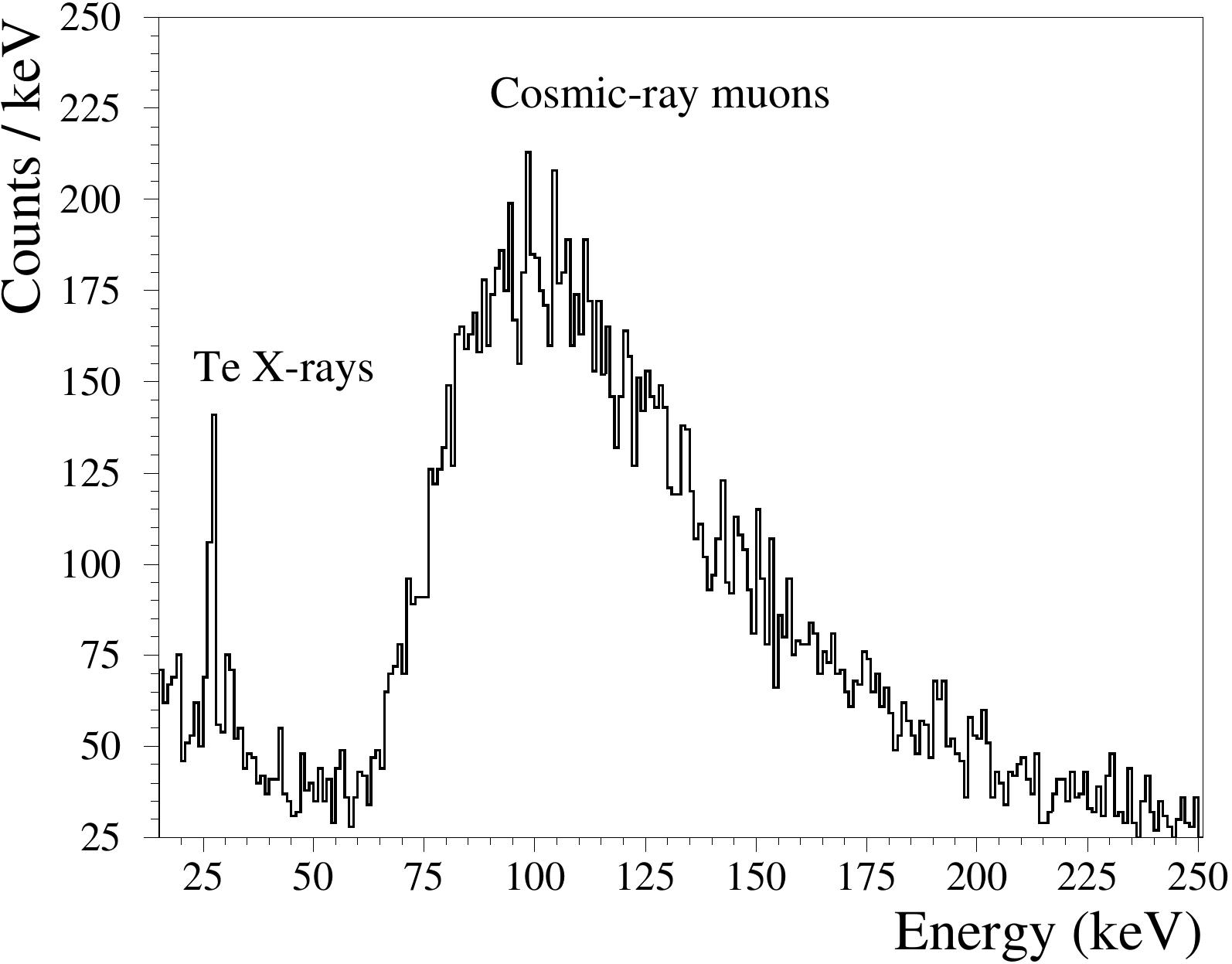}
\caption{Energy spectrum of events detected by the bolometric Ge LD over 24~h of Run I measurements in the above-ground cryogenic set-up. The spectrum exhibits Te X-ray K$_\alpha$ and K$_\beta$ peaks (fluorescence induced in the two neighboring TeO$_2$ crystals by natural radioactivity) and a cosmic-ray induced muon bump.}
\label{fig:geSpectrum}
\end{figure}

A calibration of the Ge LD in both Runs is done by using a distribution of comic-ray muons passing through the wafer, which is well-described by the Landau distribution and its maximum corresponds to the most probable muon-induced energy release. According to GEANT4-based simulations, we expect the most probable energy deposition in a 0.17-mm-thick Ge to be 100~keV~\cite{novati2019charge}. Taking into account a lack of knowledge on the precise thickness of the used Ge wafer and keeping in mind an observed spread of $\pm0.02$~mm~\cite{tenconi2015development}, an uncertainty of this calibration method in the present conditions is expected to be around 10\%. However, we were able to improve the precision of this calibration by exploiting an alternative calibration, available in Run I, to tune the value of the most probable energy of muons. The alternative calibration is realized thanks to the presence of two TeO$_2$ crystals, facing the LD on the opposite side of the ZnO sample, which can act as Te X-ray sources being exposed to radiation~\cite{berge2018complete}. Indeed, we clearly see a Te X-ray K$_\alpha$/K$_\beta$ doublet in addition to the muon distribution, as illustrated in figure~\ref{fig:geSpectrum}. Thanks to this observation, we calibrated the most probable muon-induced energy as 93~keV and used this value for the LD energy scale determination. Consequently, we found a rather high LD sensitivity of about 3~$\upmu$V/keV in Run I, which was then reduced by a factor of 3~in Run II. Despite a significant difference in the sensitivity, the LD noise was measured to be at a similar level of 0.2--0.3~keV FWHM. By applying a 60~V bias on the Al electrode, signal amplification by a factor of 12~is achieved, while the noise reduction factor is approximately 7, decreasing the LD noise to around 40~eV FWHM.

\begin{figure}[h!]
\centering
\includegraphics[width=0.7\textwidth]{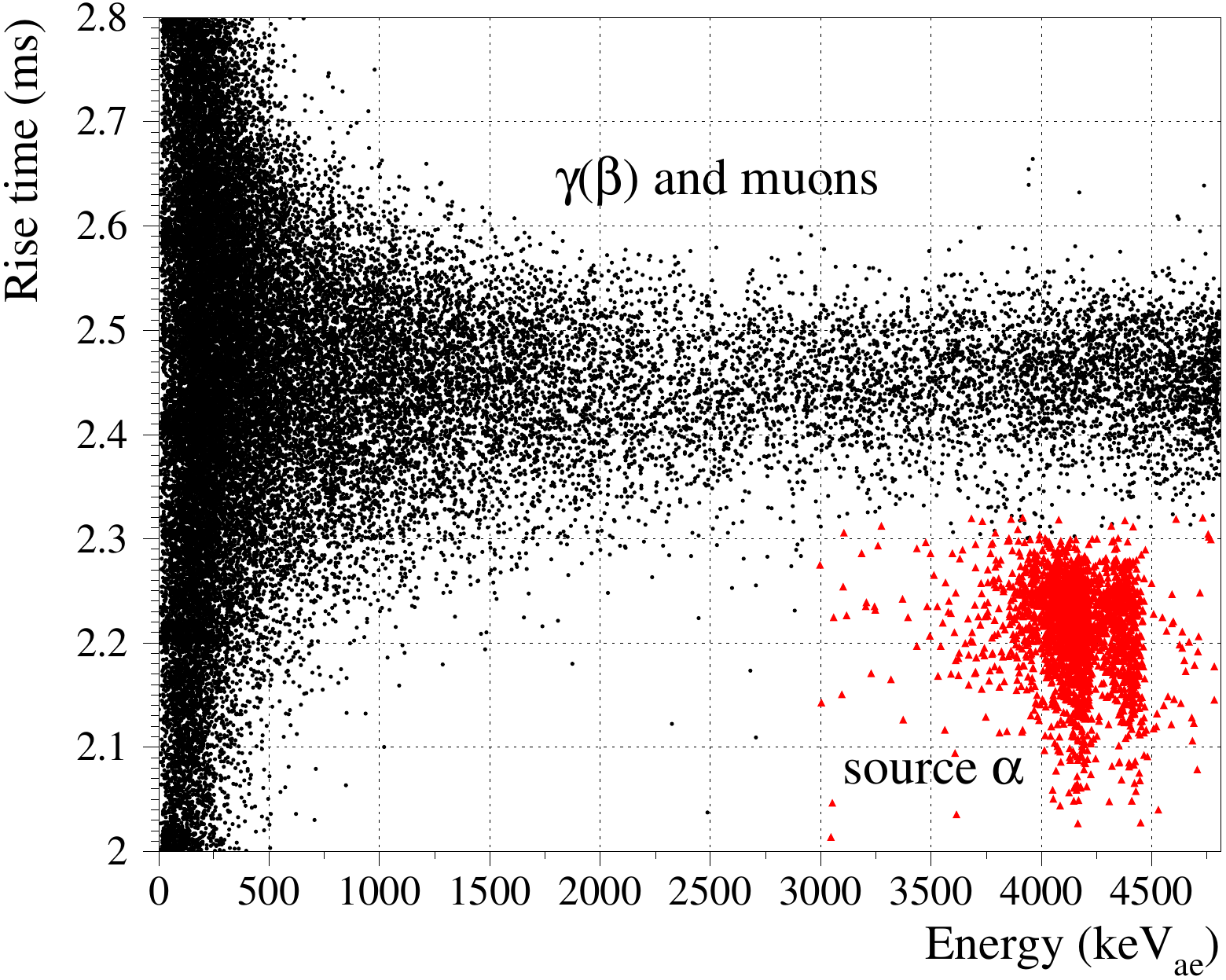}
\caption{Distribution of Rise time parameter for different type of events (black $\gamma, \beta, \upmu$'s; red $\alpha$'s) detected by the ZnO scintillating bolometer in Run I (96~h of measurements). The energy scale is calibrated using $\alpha$ particles (keV alpha equivalent, keV$_{\textrm{ae}}$). The detectors shows pulse-shape discrimination capability using heat signals, as demonstrated by separation of $\alpha$ particles (mostly originated to the source) from $\gamma,\beta$, muon-induced events.}
\label{fig:riseTime}
\end{figure}

Taking into account the results of the ZnO sample $\gamma$ screening (limits on \Nuc{Th}{228}/\Nuc{Ra}{226} activity are at the level of ten(s) mBq/kg, see table~\ref{tab:hpge}), a total $\alpha$ activity of Th/U radionuclides in the studied crystal is expected to be low (i.e.~ten(s) events per day). Thus, in Run I we decided to use an external $\alpha$ source (\Nuc{U}{238}/\Nuc{U}{234}, a smeared energy profile~\cite{armatol2021cupid}) in order to investigate the response of the ZnO bolometer to $\alpha$ particles in comparison to similar energy electron recoil signals induced by muons. Among the above-listed pulse-shape parameters, only the Decay time parameter shows a small difference between types of particles, while other parameters clearly show separate populations. An example of efficient particle identification achieved using a rather simple pulse-shape parameter such as Rise time is illustrated in figure~\ref{fig:riseTime}; similar efficiency is obtained with the PSD parameter, while the Correlation appears to be less powerful.

\begin{figure}[h!]
\centering
\includegraphics[width=0.7\textwidth]{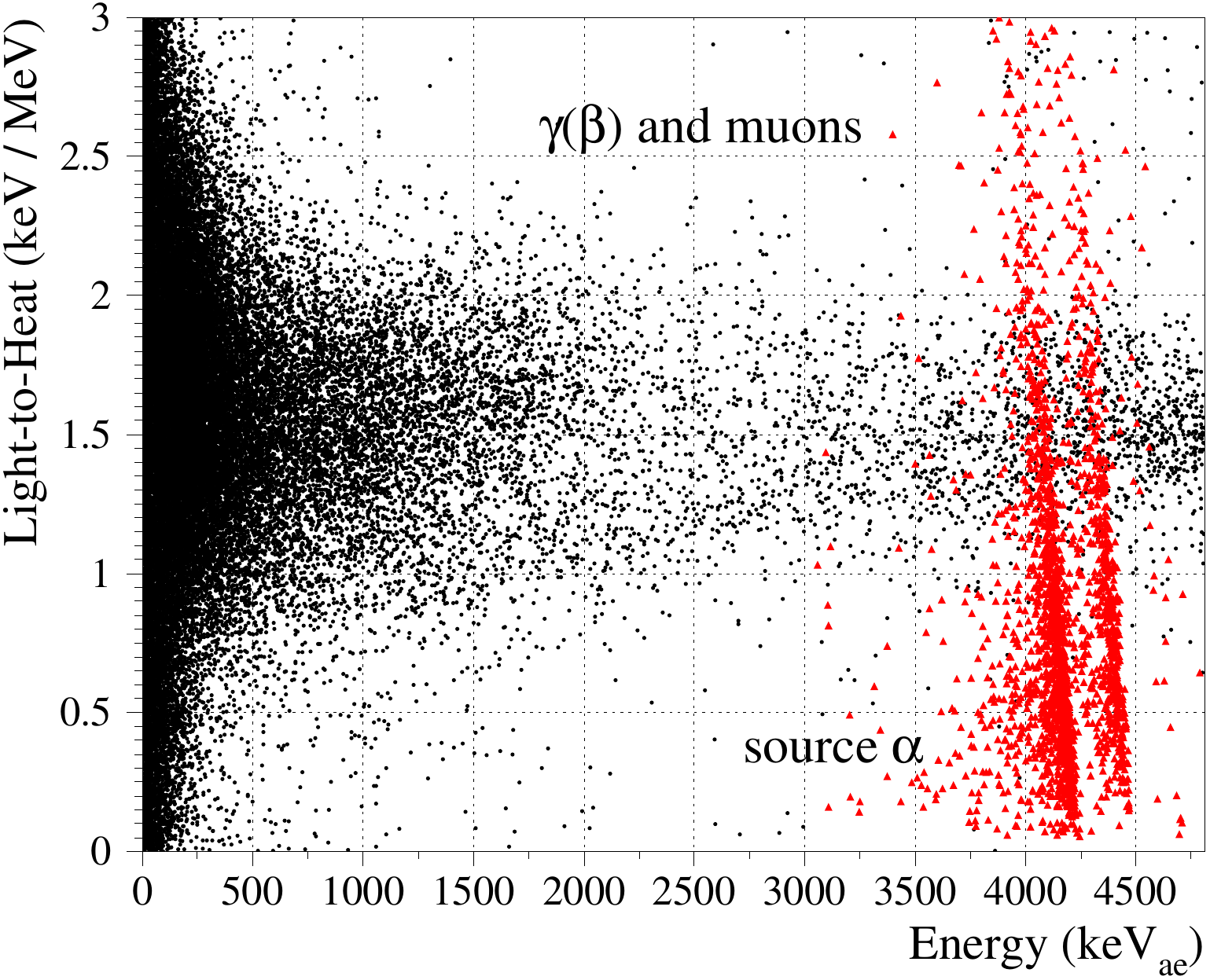}
\caption{Scatter plot of the Light-to-Heat parameter versus energy of particles detected by the ZnO scintillating bolometer in Run I (96~h of measurements). The $\alpha$ particles (red distribution) are selected from $\gamma, \beta,$ and $\upmu$'s (black distribution) using the Rise time parameter.}
\label{fig:lightToHeat}
\end{figure}

Thanks to the use of the photodetector, we can explore simultaneous detection of particle-induced energy release in the ZnO bolometer followed by scintillation detected by the LD as a viable tool of particle identification~\cite{poda2021scintillation,poda2017low}. Coincidences between ZnO and LD channels were established using ZnO-detected events as a trigger and taking into account a tiny difference in a rising part of signals of the detectors. Often, detected scintillation is given relative to the corresponding heat energy, in keV/MeV, represented by the Light-to-Heat ratio. An example of the distribution of the Light-to-Heat parameter versus energy of detected events in a dataset of Run I (the same events shown in figure~\ref{fig:riseTime}) is illustrated in figure~\ref{fig:lightToHeat}. A distribution of $\alpha$ events exhibit enhanced signals detected by the LD, leaking into the band of $\beta/\gamma/\upmu$-induced events and even further.

\begin{figure}[h!]
\centering
\includegraphics[width=0.7\textwidth]{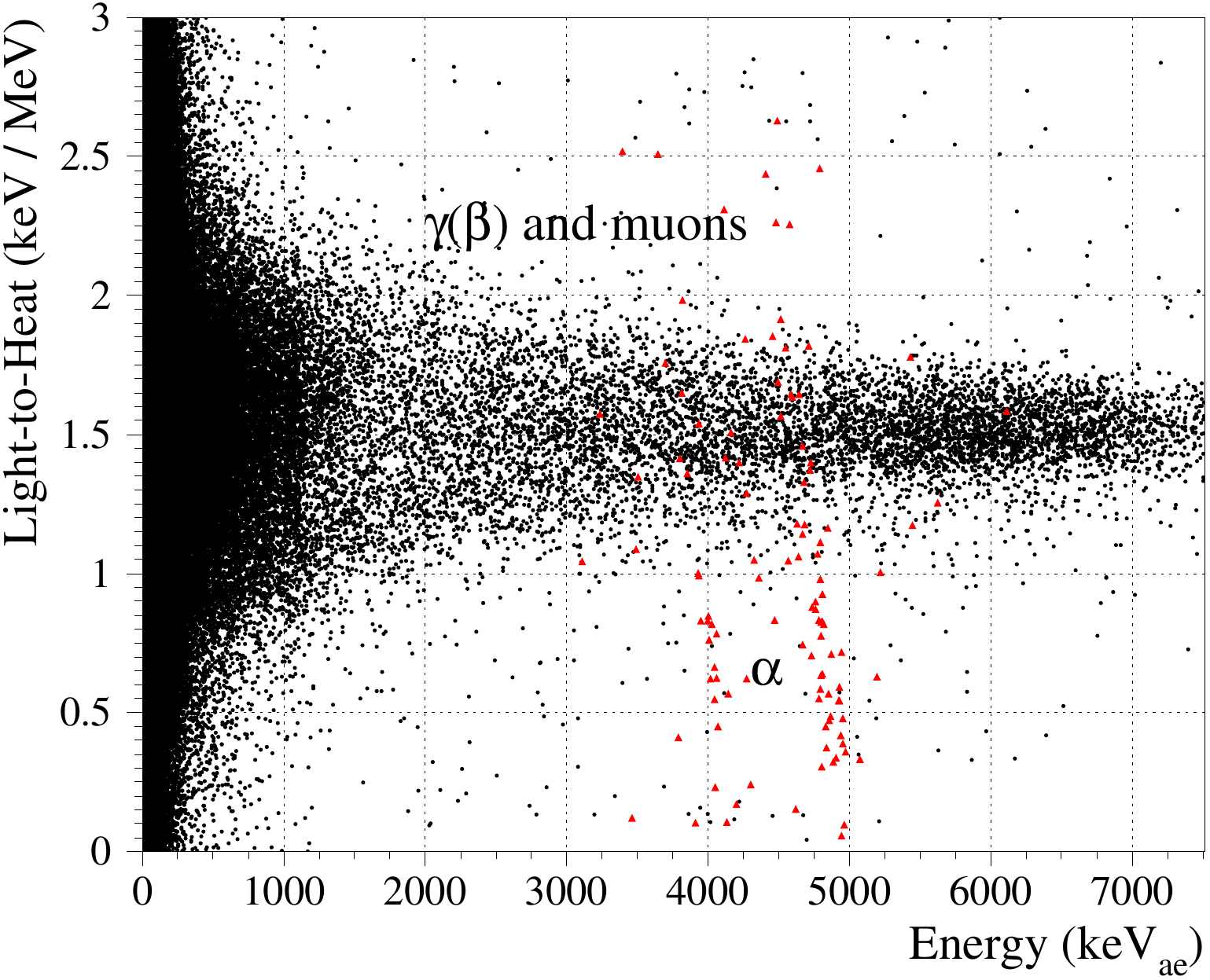}
\caption{Two-dimensional histogram showing a distribution of Light-to-Heat parameter versus energy of particles registered by the ZnO scintillating bolometer in 271~h of measurements (Run II). The $\alpha$ particles (red distribution) are selected from $\gamma, \beta$ and $\upmu$'s (black distribution) using the PSD parameter. }
\label{fig:bkgnd}
\end{figure}

Such detector's response to $\alpha$ particles is unusual. We would explain this effect by $\alpha$ particles irradiation on the defected crystal surface. Indeed, all sides of the ZnO sample, apart from the side faced to the LD, were roughened by a 400-grid sandpaper to enhance light collection caused by diffuse reflection on crystal surfaces. This surface treatment leads to a surface damaging on a certain depth, which is typically estimated to be three times of abrasive particulates dimensions, i.e.~75~$\upmu$m in our case. Therefore, $\alpha$ particles with energies less than 4.5~MeV would be fully stopped in the damaged layer. From the material properties point of view, this damaged layer could be enriched with low-lying charge traps, which then could be populated by charge carriers during the ZnO sample handling under sunlight. During an $\alpha$ particle interaction, a large number of such trapped charge carriers could be released around $\alpha$ particle track, and following to luminescence centers would enhance the scintillation light emission. Similar effect of the crystal defect structure involvements in the light yield enhancement was observed with ZnSe-based scintillating bolometers, where typically observed light yield for $\alpha$ particles is significantly higher than that of electrons or $\gamma$ quanta, and hence the quenching factor for $\alpha$ particles is larger than 1 ~\cite{nagorny2017quenching,silva2018characterization}. Background measurements, i.e.~without an externally located calibration $\alpha$ source, demonstrated in figure~\ref{fig:bkgnd}, supporting a hypothesis that the defect structure of the ZnO bulk material plays a major role to the observed effect of the light yield enhancement for internal $\alpha$ particles too.

The total internal $\alpha$ activity evaluated in the 3.0--7.5~MeV energy interval is 22(2)~mBq/kg, being calculated taking into account the selection efficiency of the ``pure'' $\alpha$ events through the pulse-shape analysis (90\%). Moreover, in the energy spectrum collected in the background run, i.e. without an external calibration $\alpha$ source, there are two prominent events distributions that could be ascribed to the ZnO crystal internal contamination by 
6(1)~mBq/kg of \Nuc{Th}{232} and 12(2)~mBq/kg of \Nuc{U}{234}. To evaluate the internal $\alpha$ contamination of the ZnO sample more precisely a longer background run is required, as well as a proper correction of the light yield enhancement for $\alpha$ particles should be taken into consideration in order to significantly improve energy resolution, in a similar manner as it was done in~\cite{arnaboldi2010cdwo4}. Nevertheless, from the current data set, one can conclude that the secular equilibrium in natural decay chains is broken, which could be an evidence of a very effective segregation effect occurred during the crystal growth.

\section{Searches for double beta decay processes in zinc
}\label{sec:dbd}

In this study we do not posses the low background conditions and large exposure required for high-sensitivity searches for DBD processes (i.e.~with currently leading half-life sensitivities $\mathcal{O}(10^{21}$--$10^{26})$ yr, depending on DBD channels and isotope of interest). This is due to the above ground location of this experiment (i.e.~significant cosmic-ray background), modest external shielding around the cryostat (i.e.~significant environmental $\gamma$-ray background), no special care on radiopurity of the cryostat and the detector component, less than 10~g mass of the ZnO sample, and a short duration of measurements (hundreds of hours). Despite the aforementioned constrains, we can use the present measurements to show prospects of ZnO bolometers for DBD search applications, particularly thanks to a low energy threshold and a good energy resolution demonstrated in this study.

\begin{figure}[h!]
\centering
\includegraphics[width=0.7\textwidth]{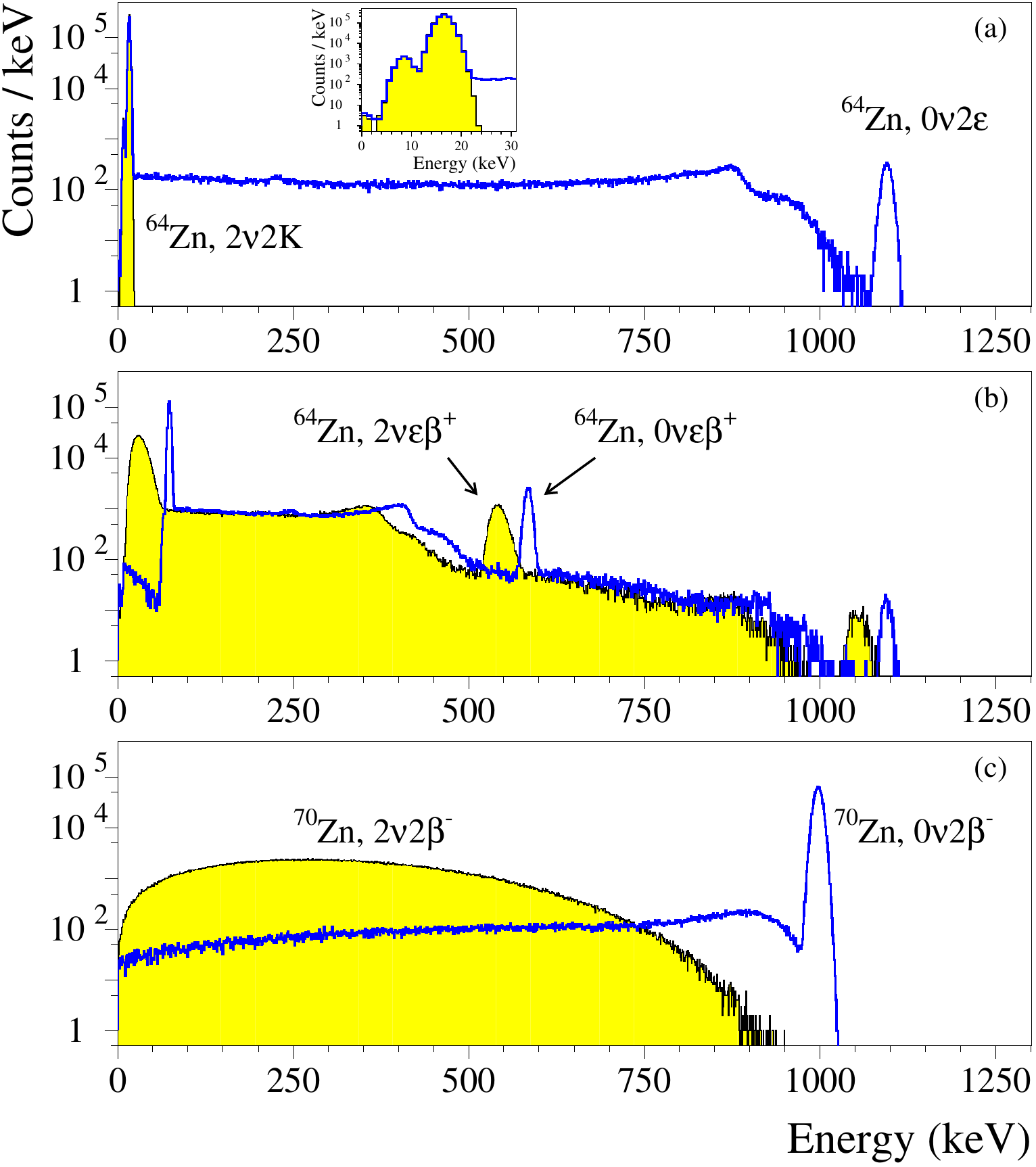}
\caption{Energy spectra of GEANT4-based MC simulations of DBD processes in the operated ZnO bolometer: $2\nu$ and $0\nu$ modes of double-electron capture (a) and electron capture with positron emission (b) in \Nuc{Zn}{64}, and double-beta decay of \Nuc{Zn}{70} (c) Each distribution corresponds to $10^6$ decays; the energy dependence of the ZnO bolometer energy resolution has been taken into account.}
\label{fig:mc1}
\end{figure}

First of all, we performed MC simulations of DBD processes in the ZnO bolometer aiming at obtaining their response functions. We used a GEANT4-based application Simourg~\cite{kobychev2011program} to construct a simplified geometry of the detector and to run MC simulations using the kinematics of $10^6$ initial particles emitted in DBD transitions, provided by the DECAY0 event generator~\cite{ponkratenko2000event}. The GEANT4 10.2 (patch 02) libraries and Livermore low-energy electromagnetic physics list have been used to run the Simourg program. The energy dependence of the ZnO energy resolution has been approximated using several $\gamma$ peaks found in the ZnO background spectrum. The resulting MC-simulated energy distributions of DBD events expected for different DBD processes in Zn isotopes embedded in the ZnO bolometer are illustrated in figure~\ref{fig:mc1}.

As it is seen in figure~\ref{fig:mc1}, all simulated response functions have at least one peak-like feature which might help to recognize a DBD signal of interest over a flat background except the featureless, continuous distribution from \Nuc{Zn}{70} $2\nu2\beta^-$ decay events. We used these features to optimize the region of interest (ROI) in terms of the DBD signal containment efficiency and the measured background (e.g. using a ratio of the detection efficiency to a square root of number of counts in the ROI). We found that the optimal ROIs to search for DBD processes in \Nuc{Zn}{64} are in the energy interval of 10--100~keV, in particular $\sim$10--20~keV for \Nuc{Zn}{64} double-electron capture (more probable process compared to $\varepsilon\beta^+$ decay modes thanks to a higher energy available; one can also compare phase space factors~\cite{kotila2013phase}). Such a comparatively low-energy threshold is routinely demonstrated by macro-bolometers including the ZnO low-temperature detector of the present work, where 2 and 6 keV energy thresholds (defined as $5~\times$~RMS baseline noise) were used in the data processing of Runs I and II respectively. In the case of \Nuc{Zn}{70} $0\nu2\beta^-$ decay, a peak (i.e.~in the ROI) is expected at 1~MeV, while for the $2\nu$ mode the highest counting rate would be at $\sim$1/3 of $Q_{2\beta}$. However, for the $2\nu$ mode we used a ROI corresponding to a higher energy part of the spectrum, $\sim$0.4--0.7~MeV, which has an order of magnitude lower background than at the maximum of the $2\nu2\beta^-$ distribution. ROIs chosen in the present study and the corresponding detection efficiencies are listed in table~\ref{tab:dbd2}.

\begin{table}[h!]
\caption{Region of interest (ROI) to search for various DBD processes in Zn isotopes, corresponding DBD signal containment efficiencies, ultimate sensitivities, and half-life limits obtained by applying two methods to the data of the ZnO bolometer (7.2~g of mass, operated over 271~h). The ultimate sensitivities and half-life limits are presented at the 90\% confidence level. In the case of $2\nu2\beta^-$ decay of \Nuc{Zn}{70}, we conservatively ascribe all events in the ROI to a continuum signature of the process searched for. } 
\begin{center}
\begin{tabular}{cccccc}
\hline
 \multirow{2}{*}{Nuclide}& DBD & ROI & DBD containment & Ultimate & Half-life \\
 & process & [keV] & efficiency, $\eta$ & sensitivity [yr] & limit [yr] \\ \hline
\multirow{4}{*}{\Nuc{Zn}{64}} & $2\nu2$K & 15--60 & 0.960 & 1.8$\times10^{18}$ & 8.8$\times10^{17}$ \\
& $0\nu2\varepsilon$ & 15--60 & 0.826 & 1.5$\times10^{18}$ & 7.5$\times10^{17}$  \\
& $2\nu\varepsilon\beta^+$ & 30--90 & 0.424 & 7.6$\times10^{17}$ & 1.0$\times10^{17}$  \\
& $0\nu\varepsilon\beta^+$ & 40--90 & 0.646 & 1.3$\times10^{18}$ & 3.6$\times10^{18}$  \\ \hline
\multirow{2}{*}{\Nuc{Zn}{70}} & $2\nu2\beta^-$ & 430--730 & 0.240 & 1.3$\times10^{14}$ & 3.5$\times10^{14}$ \\
 & $0\nu2\beta^-$  & 900--1100 & 0.914 & 6.1$\times10^{16}$ & 1.7$\times10^{17}$ \\ \hline
\end{tabular}
\end{center}
\label{tab:dbd2}
\end{table}

In order to estimate the sensitivity of the experiment to different DBD processes of interest, the half-life limits were calculated using the following formula:
\begin{equation}\label{eq:lim}
\lim T_{1/2} = \ln2 \cdot N \cdot \eta_{PSD} \cdot \eta \cdot t / \lim S,
\end{equation}
where $N$ is a number of nuclei of interest ($2.55\times10^{22}$ nuclei of \Nuc{Zn}{64} isotope and $3.16\times10^{20}$ nuclei of \Nuc{Zn}{70}), $\eta_{PSD}$ is a selection efficiency of events, $\eta$ is a DBD signal detection efficiency, $t$ is a time of measurements (271~h), and $\lim$\textit{S} is a number of counts excluded at a given confidence level. The event selection efficiency (including also a trigger efficiency) has been studied using a pulse injection method (i.e. pulse templates with different amplitudes were randomly injected into the data stream and processed in the same way as physics data to evaluate a selection efficiency at a given energy). 

\begin{figure}[h!]
\centering
\includegraphics[width=0.7\textwidth]{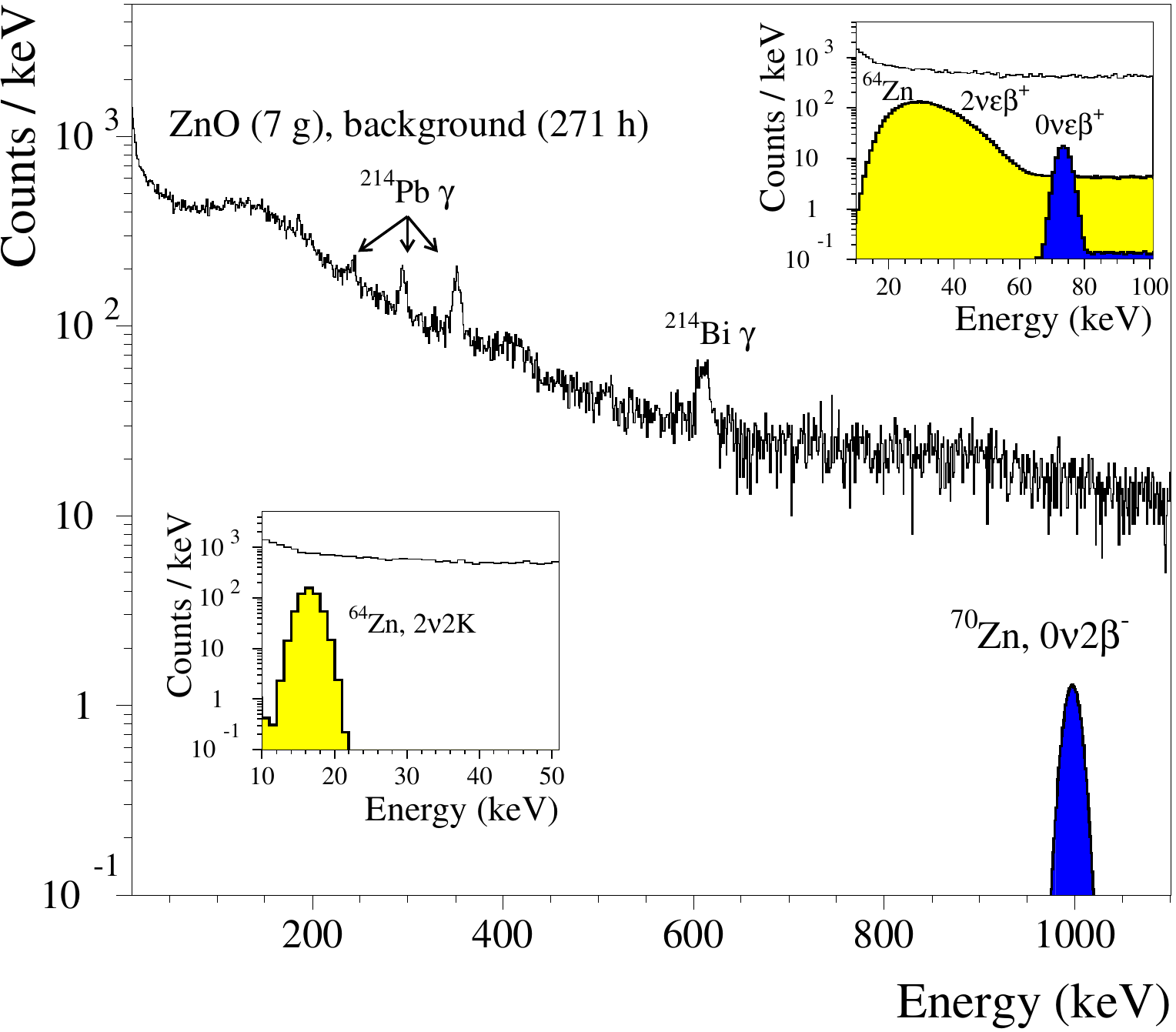}
\caption{Fragment of the energy spectrum collected with the 7.2~g ZnO bolometer in a 271-h-long measurement at sea level. The most intensive $\gamma$ peaks of environmental radioactivity present in the spectrum are labeled. Distributions of different DBD processes in Zn isotopes, excluded at 90\% C.L. level, are shown in the main plot (\Nuc{Zn}{70}) and in the two insets (both correspond to \Nuc{Zn}{64}). For clarity of the inset, we do not present the $0\nu2\varepsilon$ decay mode which has a similar energy distribution as $2\nu2$K below 20~keV.}
\label{fig:znspectrum}
\end{figure}

To derive $\lim$\textit{S}, the ZnO sample energy spectrum was fitted in the ROIs with a simple model containing GEANT4-simulated response functions of the ZnO bolometer to various DBD processes and a background component. Using the results of the fits and the Feldman-Cousins approach~\cite{feldman1998unified}, we obtained $\lim$\textit{S} values and calculated the corresponding half-life limits. For instance, the least squares fit in the 15--60 keV (30--90 keV) interval, $\chi^2$ = 73.4/46 = 1.59 (71.2/61 = 1.17), returns the area of the $2\nu2$K ($2\nu\varepsilon\beta^+$) decay effect as $425\pm72$ counts ($3711\pm605$ counts). Consequently, we calculated a Feldman-Cousins limit on the excluded events, $\lim$\textit{S} = 543 counts (4703 counts), and a corresponding half-life limit, $\lim T_{1/2} = 8.8\times10^{17}$ yr ($1.0\times10^{17}$ yr) at 90\% C.L. The results for all searched-for DBD modes are summarized in table~\ref{tab:dbd2}. Different distributions of DBD events, excluded at the current level of sensitivity, are illustrated in figure~\ref{fig:znspectrum}.

To estimate the ultimate sensitivity of the measurement in a dedicated energy interval, we used the so-called ``1$\sigma$ approach'' in which statistical uncertainties of the number of events registered in a ROI are taken as $\lim$\textit{S} (we took 1.64$\sigma$ to comply with 90\% C.L.). For instance, 25552 counts (27985 counts) were found in the energy range of 15--60 keV (30--90 keV) of the ZnO energy spectrum, which results to $\lim$\textit{S} = 262 counts (274 counts). Using these $\lim$\textit{S} values together with $N$ of \Nuc{Zn}{64} nuclei, $\eta$ of the $2\nu2$K ($2\nu\varepsilon\beta^+$) processes, $\eta_{PSD}$ = 0.88 (0.90) in the considered energy intervals, and $t$ of data taking, one can obtain $\lim T_{1/2} = 1.8\times10^{18}$ yr $(7.6\times10^{17}$ yr). The results of this method are also presented in table~\ref{tab:dbd2}. It should be noted that in case of $2\nu2\beta^-$ decay of \Nuc{Zn}{70}, we conservatively ascribe all events in the ROI to a continuum signature of the process searched for (see table~\ref{tab:dbd2}).

\begin{table}[h!]
    \caption{Sensitivities of different experiments, $\lim T_{1/2}$, to search for various DBD processes in Zn isotopes. The $\lim T_{1/2}$ results of the present search with the ZnO scintillating bolometer and other experiments quoted are given at 90\% C.L., except a few ones labeled with (*) which are at 68\% C.L. All quoted previous experiments, except studies \cite{berthelot1953physique} and \cite{norman1985improved} were realized in underground laboratories, and in dedicated low-background facilities.} 
\begin{center}
\begin{tabular}{cccccc}
\hline
\multirow{2}{*}{Nuclide} & DBD  &  \multirow{2}{*}{$\lim T_{1/2}$ [yr]} & \multirow{2}{*}{Exposure (kg$\times$d)} &\multirow{2}{*}{Detector Type} & \multirow{2}{*}{Ref.} \\ 
& process &   &  &  & \\ \hline
\multirow{19}{*}{\Nuc{Zn}{64}} & \multirow{4}{*}{$2\nu2$K}   & $1\times10^{19}$& 193 & ZnWO$_4$ scint. det. & \cite{belli2011final}\\  
 &  &  $9\times10^{17}$   & 0.08 & ZnO scint. bol. & This work  \\ 
 &  &   $6\times10^{16}$ & 0.14 & CdZnTe semicond. & \cite{kiel2003search} \\
 &  &   $8\times10^{15}$ & 0.19 & Zn + prop. chamber & \cite{berthelot1953physique} \\ \cline{2-6}
 & \multirow{5}{*}{$0\nu2\varepsilon$} &  $3\times10^{21}$& 348 & HP Zn + HPGe & \cite{bellini2021search} \\
 & &  $3\times10^{20}$& 193 & ZnWO$_4$ scint. det. & \cite{belli2011final} \\
 & &  $7\times10^{18}$& 7.85 & CdZnTe semicond. & \cite{wilson2008cobra} \\
 & &  $8\times10^{17}$   & 0.08 & ZnO scint. bol. & This work  \\ 
 & &  $7\times10^{17}$& 0.08 & ZnWO$_4$ scint. det. & \cite{danevich2005znwo4} \\ \cline{2-6}
 & \multirow{6}{*}{$2\nu\varepsilon\beta^+$}  & $3\times10^{21}$ & 348 & HP Zn + HPGe & \cite{bellini2021search} \\
 &  & $9\times10^{20}$& 193 & ZnWO$_4$ scint. det.& \cite{belli2011final} \\
 &  & $1\times10^{20}$& 7.13 & Zn + HPGe & \cite{kim2007searches} \\
 &  & $4\times10^{18}$& 0.08 & ZnWO$_4$ scint. det.& \cite{danevich2005znwo4}~ \\
 &  & $2\times10^{18}$*& 1.53 & Zn + HPGe & \cite{norman1985improved} \\ 
 &   & $1\times10^{17}$   & 0.08 & ZnO scint. bol. & This work  \\ \cline{2-6}
 & \multirow{8}{*}{$0\nu\varepsilon\beta^+$}&  $1\times10^{22}$& 4142& ZnSe scint. bol. & \cite{azzolini2020search} \\
 &  & $9\times10^{20}$& 193  & ZnWO$_4$ scint. det. & \cite{belli2011final} \\
 &  & $1\times10^{20}$& 7.13 & Zn + HPGe    & \cite{kim2007searches} \\
 &  & $4\times10^{18}$   & 0.08 & ZnO scint. bol. & This work  \\ 
 &  & $2\times10^{18}$& 0.08  & ZnWO$_4$ scint. det. & \cite{danevich2005znwo4} \\
 &  & $2\times10^{18}$*& 1.53 & Zn + HPGe & \cite{norman1985improved} \\
 &  & $1\times10^{18}$& 7.85   & CdZnTe semicond.& \cite{wilson2008cobra} \\
 &  & $3\times10^{16}$& 0.14  & CdZnTe semicond.  &  \cite{kiel2003search}  \\ \hline
\multirow{6}{*}{\Nuc{Zn}{70}} & \multirow{3}{*}{$2\nu2\beta^-$} &  $4\times10^{18}$& 193 & ZnWO$_4$ scint. det. & \cite{belli2011final}\\
 & &  $1\times10^{16}$& 0.08 & ZnWO$_4$ scint. det.  & \cite{danevich2005znwo4} \\
 & & $4\times10^{14}$   & 0.08 & ZnO scint. bol. & This work  \\ \cline{2-6}
 & \multirow{5}{*}{$0\nu2\beta^-$} &  $2\times10^{21}$& 4142 & ZnSe scint. bol. & \cite{azzolini2020search} \\
 & & $3\times10^{19}$& 193 & ZnWO$_4$ scint. det. & \cite{belli2011final}  \\
 & & $7\times10^{17}$& 0.08 &  ZnWO$_4$ scint. det. & \cite{danevich2005znwo4} \\
 & & $2\times10^{17}$   & 0.08 & ZnO scint. bol. & This work  \\ 
 & & $1\times10^{16}$& 0.53 & CdZnTe semicond. & \cite{kiel2003search} \\  \hline
\end{tabular}
\end{center}
\label{tab:sensitivity}
\end{table}

\section{Discussion}

The achieved sensitivity in terms of half-life limits for most DBD modes of \Nuc{Zn}{64} and \Nuc{Zn}{70}  isotopes,  $\mathcal{O}(10^{17}$--$10^{18}$)~yr, is still several orders of magnitude lower than the most stringent limits ranging from $10^{21}$ to $10^{22}$~yr for these nuclei. Significantly higher exposures (a factor of $10^3-10^4$) of these leading experiments, as well as their underground location and well-shielded low-background set-ups were the major factors of such enhanced sensitivity, as shown in table~\ref{tab:sensitivity}. It should be also stressed that only a dozen nuclei among potentially $2\varepsilon$, $\varepsilon\beta^+$, $2\beta^+$ active isotopes were investigated at the level of sensitivity around or more than 10$^{21}$~yr, such as \Nuc{Ar}{36}, \Nuc{Ca}{40}, \Nuc{Ni}{58}, \Nuc{Zn}{64}, \Nuc{Kr}{78}, \Nuc{Ru}{96}, \Nuc{Cd}{106}, \Nuc{Sn}{112}, \Nuc{Xe}{124}, \Nuc{Ba}{130}, \Nuc{Ba}{132}, see~\cite{blaum2020neutrinoless} and references therein.
The current limitation of the experimental sensitivity in these studies is a consequence of few reasons: a) lower energy releases in $2\varepsilon$, $\varepsilon\beta^+$, $2\beta^+$ processes in comparison to those in $2\beta^-$ decay, that complicates background suppression; b) higher expected $T_{1/2}$ values, as a result of low decay energies and phase space factors; and c) lower natural isotopic abundances of these isotopes (typically less than 1\% with only few exceptions).

Moreover, the obtained half-life limits presented here, and the stringent existing limits, are both well below of the existing theoretical calculations~\cite{domin2005neutrino,grewe2008studies} for zinc isotopes. According to these predictions the observation of $2\nu2\beta^-$ of \Nuc{Zn}{70} is expected at $\mathcal{O}$(10$^{23}$--$10^{24}$)~yr; $2\nu2\varepsilon$ of \Nuc{Zn}{64} is expected at $\mathcal{O}$(10$^{25}$--$10^{26}$)~yr; $2\nu\varepsilon\beta^+$ decay mode of \Nuc{Zn}{64} is expected at $\mathcal{O}$(10$^{31}$--$10^{35}$)~yr. Hence, the most promising DBD processes in natural zinc isotopes to be searched for at the current level of experimental sensitivity are $2\nu2K$ and $0\nu2\varepsilon$ of \Nuc{Zn}{64}, along with $2\nu2\beta^-$ and $0\nu2\beta^-$ decay modes of $^{70}$Zn in case of its enrichment.

Therefore, one needs to have a clear strategy on how to enhance the experimental sensitivity of a Zn-based detector. Fortunately, in case of zinc isotopes, further improvement of the experimental sensitivity can be achieved rather easy by increasing the ZnO detector mass, producing crystals using a high purity ZnO raw materials (no less than 99.999\%), and by performing cryogenic measurements in a well-shielded cryostat at an underground laboratory. For instance, for an experiment with 5.1~kg of ZnO crystals, which is equivalent to zinc mass in the CUPID-0 detector (i.e.~4.1~kg of Zn) measured for one year, the experimental sensitivity will increase by a factor 150 with respect to the current experiment with the 7.2~g ZnO crystal. Further, if this detector would be placed in the underground facility (e.g.~LNGS or Modane) providing effective shielding against cosmic ray induced events, this would lead to reduction of the background rate by a factor of 10$^4$~\cite{laubenstein2004underground}. Thus, in total, one could expect an enhancement in the experimental sensitivity by a factor of about 10$^6$, corresponding to half-life limits of $\mathcal{O}(10^{23}$--$10^{24}$)~yr. Such experimental sensitivity would be rather competitive comparing even to current experiments in DBD field, being achieved with natural zinc.

Moreover, there is a real opportunity to further increase the experimental sensitivity (by a factor 2) by using zinc enriched in $^{64}$Zn isotope. A water solution of ZnO and Zn-containing organic compound (Zn-acetate, Zn(CH$_{3}$COO)$_{2}$$\times$H$_{2}$O) are widely used in nuclear industry as an effective reagent to prevent bacteria cultivation in water environment of an active zone of water-water reactors and to reduce corrosion of the construction materials~\cite{nukem-website}. Zinc used for this application should be depleted in $^{64}$Zn isotope, since this isotope has the highest cross-section of neutron capture among all natural zinc isotopes leading to a notable activation under a high flux of neutrons in the active zone. Therefore, there are significant quantities of $^{64}$Zn remaining after zinc isotopes depletion for nuclear industry application, which is considered as a by-product. The cost of the highly enriched (more than 99.9\%) $^{64}$Zn isotope is $\mathcal{O}(10)$~kEuro/kg compared to the typical price of enriched isotopes used in the DBD field ($\mathcal{O}(100)$~kEuro/kg), which makes it affordable for a 10 kg-scale experiment implementation.  

One more possibility to enhance the experimental sensitivity that could be achieved with ZnO-based scintillating bolometer is related to the fact that the ZnO crystal is a semiconductor material. Therefore, one can exploit the phonon signal amplification in the presence of the electric field, i.e. the Neganov-Trofimov-Luke (NTL) effect, as it was earlier adopted to Ge light detector~\cite{novati2019charge}. This phonon signal amplification will lead to increasing of signal amplitude, followed by an improvement of the baseline energy resolution and significant lowering of the low-energy detection threshold. For instance, typical energy resolution for cryogenic light detectors $\mathcal{O}(300)$~eV FWHM~\cite{beeman2013characterization} can be improved to $\mathcal{O}(50)$~eV using a NTL-amplification~\cite{novati2019charge}, as also shown in the present work. The improved energy resolution would help to minimize background contribution to the region of interest leading to an enhancement of the experimental sensitivity by a factor of $5$. All these parameters are vital for studies of DBD processes with low energy release in the outgoing channel, such as $2\nu2K$ in \Nuc{Zn}{64}. 

An important consideration that should be also taken into account during the ZnO crystal production for such experiment, their processing, and the final experiment installation, is that raw ZnO material and the produced ZnO crystals should be transported only by a land-transportation to avoid its cosmogenic activation and long-lived \Nuc{Zn}{65} isotope production that would results in experimental sensitivity reduction.

\section{Conclusion}

In this work we report on a detailed study of the performance of a ZnO-based scintillating bolometer as a detector to search for rare processes in zinc isotopes. For the first time, a 7.2~g ZnO crystal containing more than 80\% of zinc in its mass, was successfully operated as a low-temperature detector over 271 h of background measurements in a pulse-tube dilution refrigerator at a surface lab.

The ZnO-based detector exhibits the excellent energy resolution of baseline noise 1.0--2.7~keV FWHM at various working temperatures that results in a low-energy threshold of the experiment, i.e. 2.0--6.0~keV. This also makes it feasible to study rare processes with low-energy releases, like $2\nu2K$ decay mode of \Nuc{Zn}{64}.

The light yield for $\beta$/$\gamma$ events was determined to be 1.5(3)~keV/MeV, while it varies for $\alpha$ particles in the range of 0.2--3.0~keV/MeV, most probably, due to contribution of the bulk material defect structure to the light yield enhancement effect. 
The ZnO-based detector shows a powerful pulse-shape discrimination capability using time-properties of only heat signals (namely, Rise time parameter), demonstrating a full separation of alpha particles from $\beta$/$\gamma$ of environmental radioactivity and muon-induced events. 

The ZnO crystal was found to be rather radiopure with respect to daughter nuclides from the U/Th natural decay chains, as well as \Nuc{K}{40} from natural radioactivity or \Nuc{Cs}{137} of anthropogenic origins, with only limits on their activities being set at the level of $\mathcal{O}$(1--100)~mBq/kg in measurements with ultra-low-background HPGe $\gamma$-spectrometer. A 22(2)~mBq/kg of the total internal $\alpha$-activity in the 3.0--7.5~MeV energy interval was calculated analysing the ``pure'' $\alpha$ events selected from Run II data set (background measurements). Two prominent alpha events distributions in the background spectrum could be ascribed to the crystal internal contamination by 6(1)~mBq/kg of \Nuc{Th}{232} and 12(1)~mBq/kg of \Nuc{U}{234}. Further, more precise, evaluation of the ZnO crystal internal $\alpha$ contamination is required. At the same time, being produced from low chemical purity grade raw materials (99.98\%), ZnO demonstrates a feasibility of its radiopurity improvement by utilizing high purity chemicals and raw materials.

Taking into account the excellent performance of the ZnO crystal as a scintillating bolometer and profiting from 271~h of acquired background data (without $\gamma$ and $\alpha$ calibration sources), limits on DBD processes in \Nuc{Zn}{64} and \Nuc{Zn}{70} isotopes were set on the level of $\mathcal{O}(10^{17}$--$10^{18})$~yr for various decay modes. The achieved sensitivity was analyzed and compared with all previous low-background long-term experiments searching for DBD processes in zinc isotopes.

To summarize, there is a good potential for ZnO-based scintillating bolometers to search for DBD processes of Zn isotopes, especially DBD modes in \Nuc{Zn}{64}, with the most prominent spectral features at $\sim$10--20~keV. Observation of this process would require a ``source~$=$~detector'' technology with a low-energy threshold, good energy resolution, and low background, all of which are achievable with ZnO-based scintillating bolometers. Together with a high natural isotopic abundance and the potential for further isotopic enrichment, the sensitivity to DBD processes in \Nuc{Zn}{64} with a 10~kg-scale experiment is at the level of $\mathcal{O}(10^{24})$ yr. Furthermore, a ZnO-based cryogenic experiment would provide complementarity to the current DBD searches and while allowing for studies of $2\varepsilon, \varepsilon\beta^+$, and $2\beta^-$ decay modes of natural zinc isotopes at the $\mathcal{O}(10^{24})$ year level and beyond.

\section{Acknowledgements}

BB is supported by the Natural Sciences and Engineering Research Council of Canada (NSERC). SSN is supported by the Arthur B. McDonald Canadian Astroparticle Physics Research Institute. We also appreciate the fruitful collaboration with Dr.~Vladimir Lutin and grateful for the provided ZnO sample used in these studies. 
The cryostat used for the low-temperature tests here described ---installed at ICLab (Orsay, France)--- was donated by the Dipartimento di Scienza e Alta Tecnologia of the Insubria University (Como, Italy). The bolometric measurements were realized within the BINGO project, which has received funds from the European Research Council (ERC) under the European Union's Horizon 2020 research and innovation program (grant agreement No. 865844 - BINGO).



\end{document}